\documentclass{article}

\usepackage{graphicx}
\usepackage{bbm}

\newcommand{\no}{\noindent}
\newcommand{\be}{\begin{eqnarray}}
\newcommand{\ee}{\end{eqnarray}}
\newcommand{\hk}{\hspace{0.1cm}}

\newcommand{\rk}{\right)}
\newcommand{\lk}{\left(}

\begin{document}

\begin{center}
{\LARGE \bf 
The Yang-Mills vacuum in 

Coulomb gauge\footnote{Invited talk given by H.
Reinhardt at the Confinement conference, Sardinia 2004}}

\vspace{5mm}

{\bf
D. Epple, C. Feuchter, H. Reinhardt}

\vspace{5mm}

Auf der Morgenstelle 14 \\
72076 T\"ubingen \\
Germany
    
\end{center}

\begin{abstract}
The Yang-Mills Schr\"odinger equation is solved in Coulomb gauge for the vacuum
by the variational principle using an ansatz for the wave functional, which is
strongly peaked at the Gribov horizon. We find an infrared suppressed gluon propagator, an 
infrared singular ghost propagator and an almost linearly rising confinement potential.
Using these solutions we calculate the eletric field of static color charge distributions
relevant for mesons and baryons.
\end{abstract}

\date{\today}

\vspace{10mm}

\no
We report on the variational solution of the Yang-Mills Schr\"odinger equation
in Coulomb gauge $\partial_i A_i = 0$ performed in \cite{R9a}. In this gauge the Yang-Mills
Hamiltonian reads
\be
\label{1}
H & = & \frac{1}{2} \int J^{- 1}[A] \Pi J[A] \Pi + \frac{1}{2} \int B[A]^2
+ \frac{g^2}{2} \int \rho (\hat{D} \partial)^{- 1} (- \partial^2) (-
\hat{D} \partial)^{-1} \rho \hk .
\ee
Here $J [A] = \det (- \hat{D}[A] \partial)$ is the Faddeev-Popov determinant, $B [A]$ is the
color magnetic field, $\Pi (x) = \delta/i \delta A(x)$ 
is the momentum operator representing the color electric field and 
$\rho (x) = - \hat{A}(x) \Pi(x)$ is the non-Abelian color charge. 
We use the following ansatz for the Yang-Mills wave functional
\be
\label{2}
\Psi = {\cal N} J^{- \frac{1}{2}} [A] \exp \lk - \frac{1}{2} \int d^3 x d^3 x' A(x) \omega (x-x') A(x') \rk 
\hk ,
\ee
where the kernel $\omega (x - x')$  is determined by minimizing
the vacuum energy
\be
\label{3}
\langle \Psi | H | \Psi \rangle = \int D A J[A] \Psi^* [A] H \Psi [A] \hk. 
\ee
Thereby we restrict ourselves to 2-loop diagrams. Minimization of the energy
gives rise to a set of coupled Schwinger-Dyson equations for the gluon
propagator
\be
\label{4}
\langle \Psi | A^a_i (x) A^b_j (x') | \Psi \rangle = \frac{1}{2} \delta^{a b}
t_{i j} (x) \omega^{- 1} (x - x')   \hk ,
\ee
the ghost form factor $d$ defined by
\be
\label{5}
\langle \Psi | (- \hat{D} \partial)^ {- 1} | \Psi \rangle = \frac{d}{- \partial^2}  \hk ,
\ee
the Coulomb form factor 
\be
\label{6}
\langle \Psi | (- \hat{D} \partial)^ {- 1} (- \partial^2) (- \hat{D} \partial)^{- 1} | \Psi
\rangle = \frac{d^2 f}{- \partial^2}  
\hk 
\ee
and the curvature in the space of gauge orbits
\be
\label{7}
\chi = - \frac{1}{2} \frac{\delta^2 \ln J [A]}{\delta A \delta A} \hk .
\ee

Resorting to the angular approximation the Schwinger-Dyson equations can be
solved analytically in the infrared $k \to 0$
\be
\label{8}
\omega (k) = \chi (k) \sim \frac{1}{k}, \hspace{0.5cm}
d (k) \sim \frac{1}{k} \hk ,  \hspace{0.5cm} \hk f (k) \to \mbox{const}  \hk 
\ee
and in the ultraviolet $k \to \infty$ 
\be
\label{9}
\omega (k) \to k , \hk \hspace{0.5cm} \frac{\chi (k)}{\omega (k)} \to 1/\sqrt{\ln k} ,
\hspace{0.5cm} d (k) \sim 1/\sqrt{\ln k} , \hk \hspace{0.5cm} f (k) \sim 1/\sqrt{\ln k}  \hk .
\ee
Here the ghost form factor has been assumed to fullfil the so-called horizon
condition $d (k \to 0) \to \infty$, but otherwise the above aymptotic behaviour
is independent of the details of the renormalization. The numerical results are
shown in figure 1. The gluon energy $\omega (k)$ is infrared divergent
signalling gluon confinement.
\begin{figure}
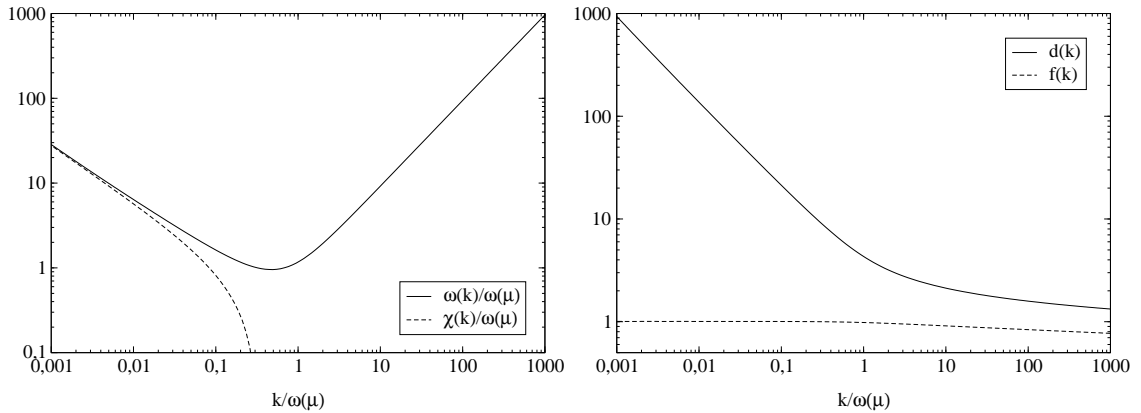

\includegraphics[scale=0.38,bb=47 33 602 436,clip=]{figure_2b}
\includegraphics[scale=0.38,bb=47 33 602 436,clip=]{figure_3}
\caption{\label{fig1} Solution for the gap function ${\omega} (k) $ (left) and Ghost form function $d (k)$ with 
Coulomb correction $f (k)$ (right). }
\end{figure}
\begin{figure}
\includegraphics[angle=90,scale=0.6]{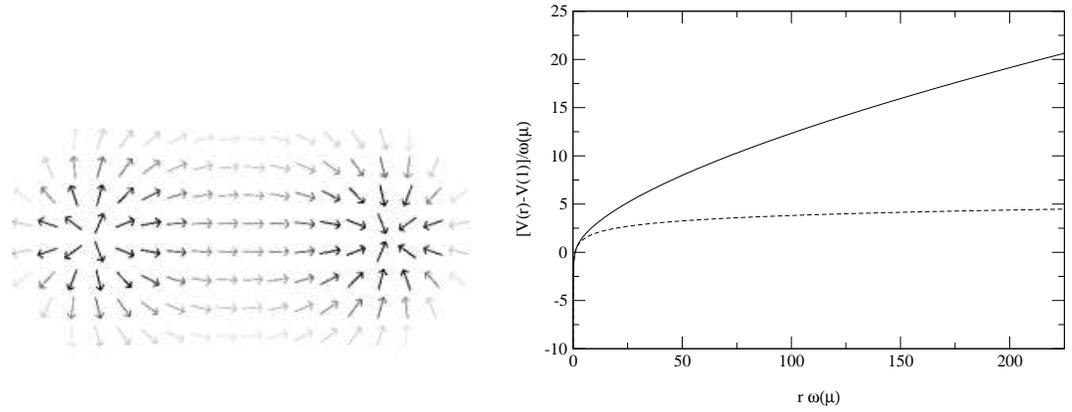}
\includegraphics[scale=0.38,bb=32 32 591 436,clip=]{figure_9d}
\caption{\label{fig4} (left) Field lines of the longitudinal chromelectric field of a charge-anticharge pair.\newline (right) Coulomb Potential with (full line) and without
inclusion of the curvature (dashed line).}
\end{figure}
\begin{figure}
\includegraphics[bb=0 0 265 335,angle=90,scale=0.5]{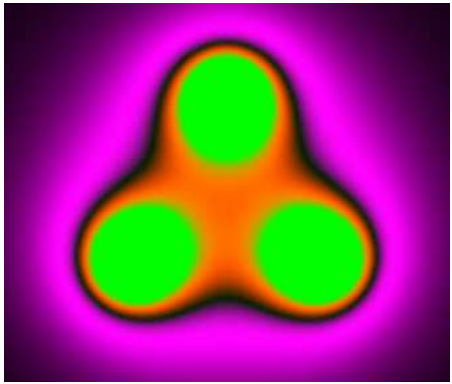}
\caption{\label{fig5} The magnitude of the longitudinal chromoelectric field of a
three-quark color-singlet state.}
\end{figure}

The electric field generated by a static charge distribution $\rho (x)$ is
given by
\be
\label{10}
E (x) = - \partial_x \int d^3 x' \langle \Psi | \langle x | (- \hat{D} \partial)^{-
1} | x' \rangle | \Psi \rangle \rho (x') \hk .
\ee
In figure 2 we show the electric field generated by a static quark-antiquark
pair. One observes the formation of a color flux tube between the static charges
and accordingly we find an (almost) linearly rising confinement potential, see
figure 2. Figure 3 shows the module of the static electric field induced by three static color
charges in a color singlet state. The flux distribution seems to prefer a
so-called Y-shape. Let us also mention that similar calculations, however,
with a different ansatz for the wave functional and ignoring the curvature fully
\cite{R16} or partly \cite{R17}, have been carried out.

Recently, we have been able to show, that the above presented results do not
depend on the detailed ansatz for the wave functional, but does crucially
depend on the curvature $\chi$ in the space of gauge orbits \cite{R100}. In particular, the
infrared limit is uniquely determined and to 1-loop order the vacuum wave
functional becomes $\Psi [A] = 1$.

The approach presented above is rather encouraging and calls for a more detailed
study of the vacuum properties of Yang-Mills theory.


\begin{thebibliography}{99}
\bibitem{R9a}
C. Feuchter and H. Reinhardt, Phys. Rev. {\bf D70}, (2004) 105021, hep-th/0408236
\bibitem{R16}
A.P. Szczepaniak and E.S. Swanson, Phys. Rev. {\bf D65} (2002) 025012
\bibitem{R17}
A. P. Szczepaniak, Phys. Rev. {\bf D69}, (2004) 074031 
\bibitem{R100}
H. Reinhardt and C. Feuchter, On the Yang-Mills wave functional in Coulomb gauge, hep-th/0408237
\end{thebibliography}
\end{document}